\documentclass[preprint,amsmath,amssymb]{revtex4-2}

\usepackage{graphicx}
\usepackage{dcolumn}
\usepackage{bm}
\usepackage{float}

\begin{document}


\title{Quantifying the effects of quarantine using an IBM SEIR model on scalefree networks}


\author{Vitor M. Marquioni}
\email{vimarqmon@gmail.com}
\author{Marcus A.M. de Aguiar}
\email{aguiar@ifi.unicamp.br}
\affiliation{Instituto de F\'isica `Gleb Wataghin', Universidade Estadual de Campinas - 13083-859, Campinas, SP, Brazil}

\date{\today}

\begin{abstract}
The COVID-19 pandemic led several countries to resort to social distancing, the only known way to slow down the spread of the virus and keep the health system under control. Here we use an individual based model (IBM) to study how the duration, start date and intensity of quarantine affect the height and position of the peak of the infection curve. We show that stochastic effects, inherent to the model dynamics, lead to variable outcomes for the same set of parameters, making it crucial to compute the probability of each result. To simplify the analysis we divide the outcomes in only two categories, that we call {\it best} and {\it worst} scenarios. Although long and intense quarantine is the best way to end the epidemic, it is very hard to implement in practice. Here we show that relatively short and intense quarantine periods can also be very effective in flattening the infection curve and even killing the virus, but the likelihood of such outcomes are low. Long quarantines of relatively low intensity, on the other hand, can delay the infection peak and reduce its size considerably with more than $50\%$ probability, being a more effective policy than complete lockdown for short periods.

\end{abstract}

\maketitle


\section{\label{sec:introduction}Introduction}

The novel Coronavirus \cite{zhu2020} pandemic has changed the lives of millions of people around the world. The lack of effective medications or a vaccine,\cite{who2020,sanders2020,keener2020,lurie2020} has made social distancing the only reliable way to slow down the virus transmission and prevent the collapse of the health system.\cite{ferguson2020,walker2020,peak2017,cyranoski2020} Quarantine measures are, however, difficult to implement and have enormous economic consequences. This is leading many communities, from cities to entire countries, to end quarantine even before the infection curve has reached its peak.\cite{erdbrink2020,bbc12020,bbc22020} Understanding how quarantine duration, effectiveness and starting time affect the infection curve is, therefore, key to guide public policies. 

Several approaches have been recently proposed to model the COVID-19 epidemic,\cite{cobey2004} including muti-layer networks,\cite{scabini2020} the Richards growth model,\cite{vasconcelos2020} and many others.\cite{flaxman2020,crokidakis2020,zhao2020,hellewell2020} Since our interest here is to quantify the effectiveness of a non-pharmacological intervention, we opted for a SEIR (susceptible, exposed, infected and recovered) individual based model (IBM) and studied how different types of quarantine change the infection dynamics.
Individuals are modeled as nodes of a scale-free (Barab\'asi-Albert) network \cite{barabasi1999} that can only infect their connected neighbors. Because the dynamics is stochastic, independent simulations with the same set of parameters can lead to quite different outcomes. Here we group the outcomes in only two categories, that we call {\it best} and {\it worst} scenarios. Stochasticity, a reality of the Sars-Cov-2 infection, is not captured by the mean field approximation of the SEIR model,\cite{hethcote2006} where outcomes depend deterministically on the model parameters.

We find three types of quarantine that can be effective against the epidemic: (i) relatively long (10 weeks) and intense (more than $80\%$ isolation); (ii) short (8 weeks) and of intermediate intensity (around $70\%$ isolation) and; (iii) long (12 weeks or more) with relatively low intensity ($40\%$ isolation). The first type, which completely ends the epidemic, is clearly the best but also the most difficult to achieve. The second type is feasible, but we find that most of the times (in most of the simulations) they result in worst case scenarios. The third type emerges as the most practical and easy to apply. It is not so effective as the previous types, but does decrease the infection peak by half. Also, it falls into the best case scenario more than $50\%$ of the times and even in the worst scenarios the infection peak decreases.

\section{\label{sec:model}The Model}

We model the spread of the virus using an extension of the SEIR model (\emph{susceptible}, \emph{exposed}, \emph{infected} and \emph{recovered} (or removed)). Exposed individuals simulate the incubation period of the disease, when infected subjects cannot yet pass on the virus. The mean field version of SEIR model is described by the equations
\begin{equation}
\begin{array}{ll}
\dot{S} & =  -\beta SI/N  \\
\dot{E} & =  \beta SI/N -\sigma E  \\
\dot{I} & =  \sigma E -\gamma I \\
\dot{R} & =  \gamma I 
\end{array}
\label{seir}
\end{equation}
where $N$ is the population size, $\beta$ is the infection rate, $\sigma$ the rate at which exposed become infected and $\gamma$ the recovery rate. The basic reproductive number, $R_0 = \beta/\gamma$, gives the number of secondary infections generated by the first infectious individual over the full course of the epidemic in a fully susceptible population.

\begin{figure}[th]
\begin{center}
\includegraphics[width=12cm]{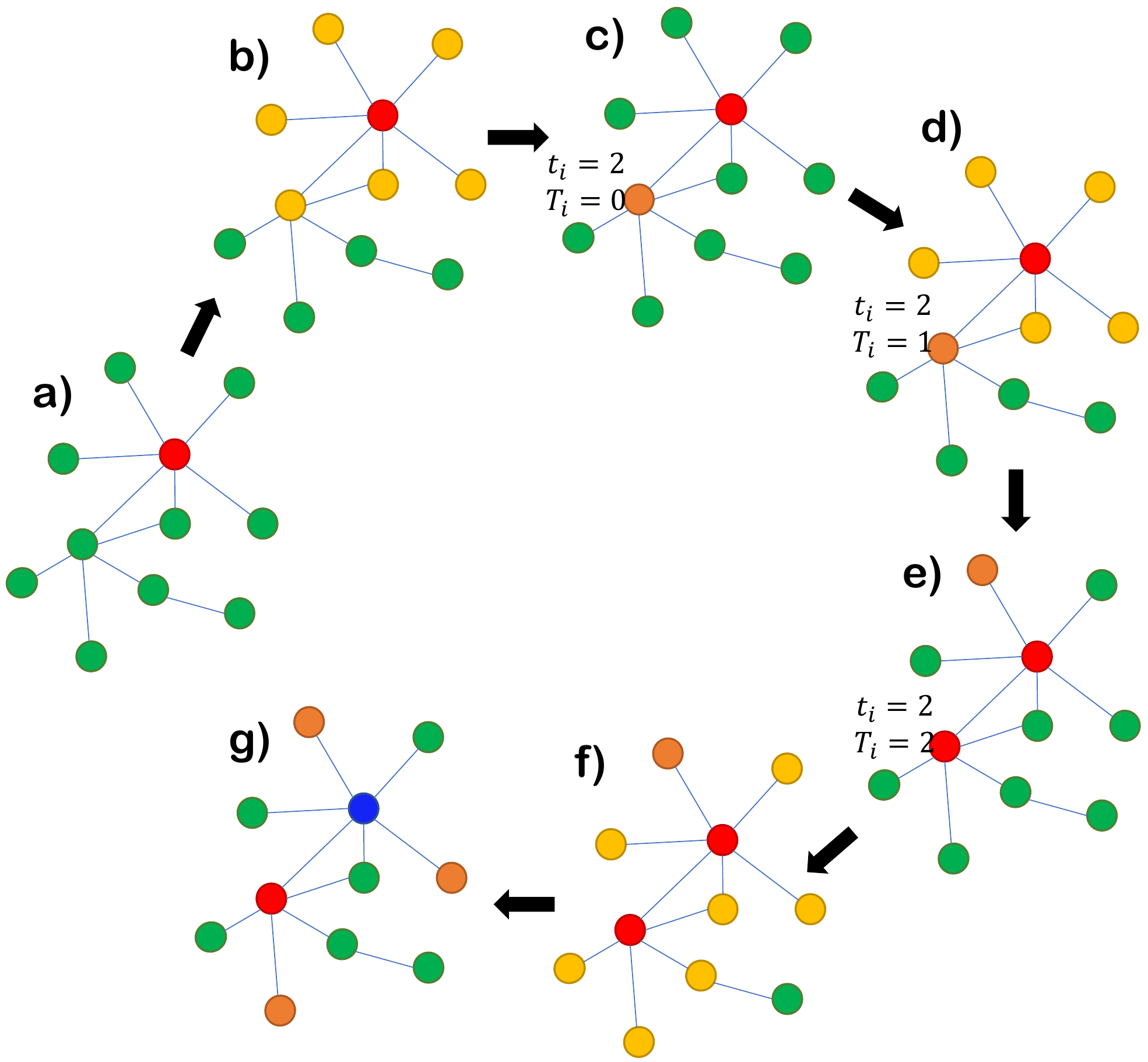}
\end{center}
\caption{Illustration of virus spread on the network: (a) initially only one individual is infected (red) and all the others susceptible (green); (b) neighbors of infected might get the virus (yellow); (c) one neighbor did get the virus and become exposed (orange); (d) neighbors of first infected individual might still get the virus, but the orange node is still in incubation time; (e) another node gets the virus from the first infected becoming exposed and the old exposed becomes infected; (f) more nodes might get the virus (yellow) and; (g) some do become exposed while the first infected becomes recovered.}
\label{fig:model}
\end{figure}

In order to take into account heterogeneity in the number of contacts we use instead an individual based model where the population is represented by a Barab\'asi-Albert network \cite{barabasi1999,albert2002,eubank2004} with $N$ of nodes and an average degree $D$. As in a deterministic SEIR model, the nodes can be classified as \emph{susceptible}, \emph{exposed}, \emph{infected} and \emph{recovered}. Susceptible individuals can become exposed if connected to an infected one; exposed individual $i$ becomes infected after a period $t_i$ of virus incubation; infected individuals can recover, and once recovered it is considered immune and therefore cannot be infected again. 

At the beginning of the simulation, only one node, chosen at random, is infected while all the remaining population is susceptible. Every susceptible node connected to the infected individual becomes exposed with the transmission probability $p$ whereas the infected node might itself recover with probability $r$.  The probability $p$ can be calculated as $p=R_0/(\tau_{symp}D)$, where $\tau_{symp}$ is the average time duration of symptoms. We assume that the symptoms last for a time $\tau$ distributed according to an exponential distribution  $\mathcal{F}(\tau)=\lambda e^{-\lambda\tau}$. Once a node is exposed, it stays exposed for an incubation time $t_i$, chosen according to a given distribution $\mathcal{P}(t_i)$.\cite{backer2020} After this period it becomes infected and is able to infect other nodes. It follows that $r\approx\lambda=1/\tau_{symp}$, for small $\lambda$. For $\mathcal{P}(t_i)$ we have used a $\Gamma(\alpha,\beta)$ distribution, as in \cite{wu2020}. Fig. \ref{fig:model} illustrates the network of contacts, states of individuals (\emph{S}, \emph{I}, \emph{E}, or \emph{R}) and the infection dynamics. For the simulations, we fixed the number of individuals $N=2000$, the average degree $D\approx98$, mean incubation time $6.5$ days with standard deviation $2.6$ days ($\alpha=6.25$ and $\beta\approx0.96$), $R_0=2.4$ and $\tau_{symp}=14$ days.\cite{cheng2020,morris2020,Linton2020} We run the simulations until the epidemic ends and no new infection is possible.

We model quarantine periods by reducing the transmission probability $p$ by the factor $(1-Q)$, where $Q$ represents the intensity of quarantine, varying from $0$ (no quarantine) to $1$ (full quarantine). Duration and starting time are specified by period $\left[t_s,t_s + t_d\right]$. We shall see that all three quarantine parameters (starting time $t_s$, duration $t_d$ and intensity $Q$) have significant effects on the dynamics, particularly on the infection peak height and delay. 

\section{\label{sec:results}Results and Discussion}

\begin{figure}[th]
\begin{center}
\includegraphics[width=16cm]{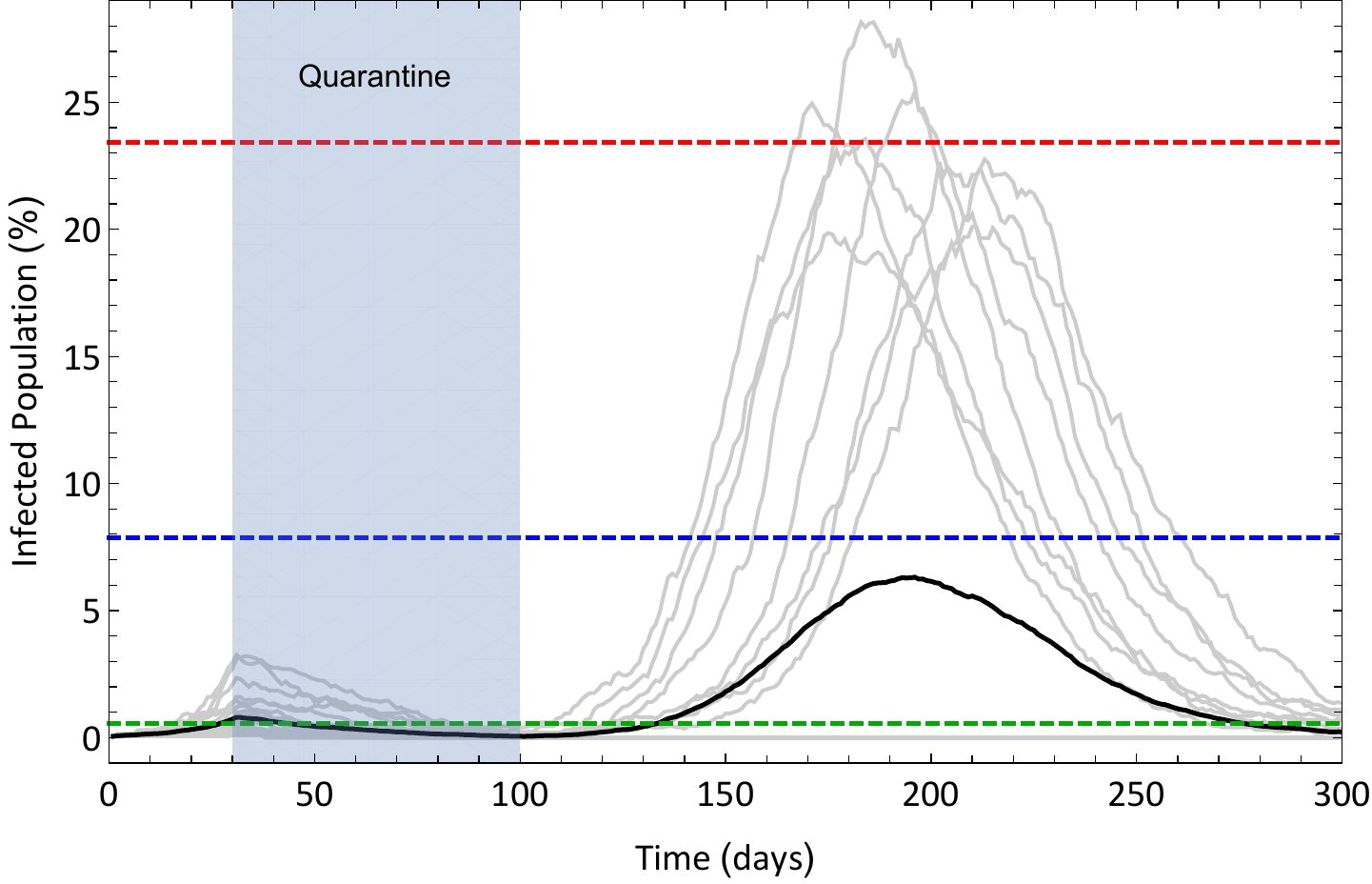}
\end{center}
\caption{Evolution of number of infected plus exposed individuals for $Q=0.9$, $t_s=30$ days and $t_d=10$ weeks for 25 replicas of the simulation. The blue dashed line shows the average height of the highest peak of each curve. Red and green dashed lines show the average peak of worst (8 replicas) and best (17 replicas) scenarios respectively, i. e., the average peak of the curves in which there is a second peak, after the quarantine, and the average peak of those in which there is not a second peak. The average of all curves (black thick line) is not representative of any actual curve. The gray shaded area indicates the quarantine period.}
\label{fig:bestworst}
\end{figure}

Unlike the mean field SEIR model, Eqs.(\ref{seir}), the present IBM version on networks is probabilistic and different outcomes are obtained every time the model is ran with the same set of parameters. To obtain statistically significant data (while keeping simulation time reasonable) we have ran the model $25$ times for different quarantine duration and intensities, beginning $t_s=20$, $30$, and $40$ days after the first infected node appears (at the beginning of the simulation). The results were divided in two different scenarios, the \emph{best} and the \emph{worst} cases. For each set of parameters, the best scenario consists of simulations where the infection peak is lower than the average peak of the full set of simulations, whereas the worst scenario contains the set with higher than average peaks. This approach is important because in many cases the epidemic response to the quarantine is not satisfactory, and this might be solely due to stochastic effects, a common feature of real systems. As an example, Fig. \ref{fig:bestworst} shows the evolution curves of infected plus exposed individuals for all 25 replicas for $Q=0.9$ and $t_d=10$ weeks. Since independent populations, represented by different Barab\'asi-Albert networks generated with the same specifications, under the same quarantine parameters might respond drastically different to quarantine, we also need to know the probability of each outcome.

\begin{figure}[th]
\begin{center}
\includegraphics[width=16cm]{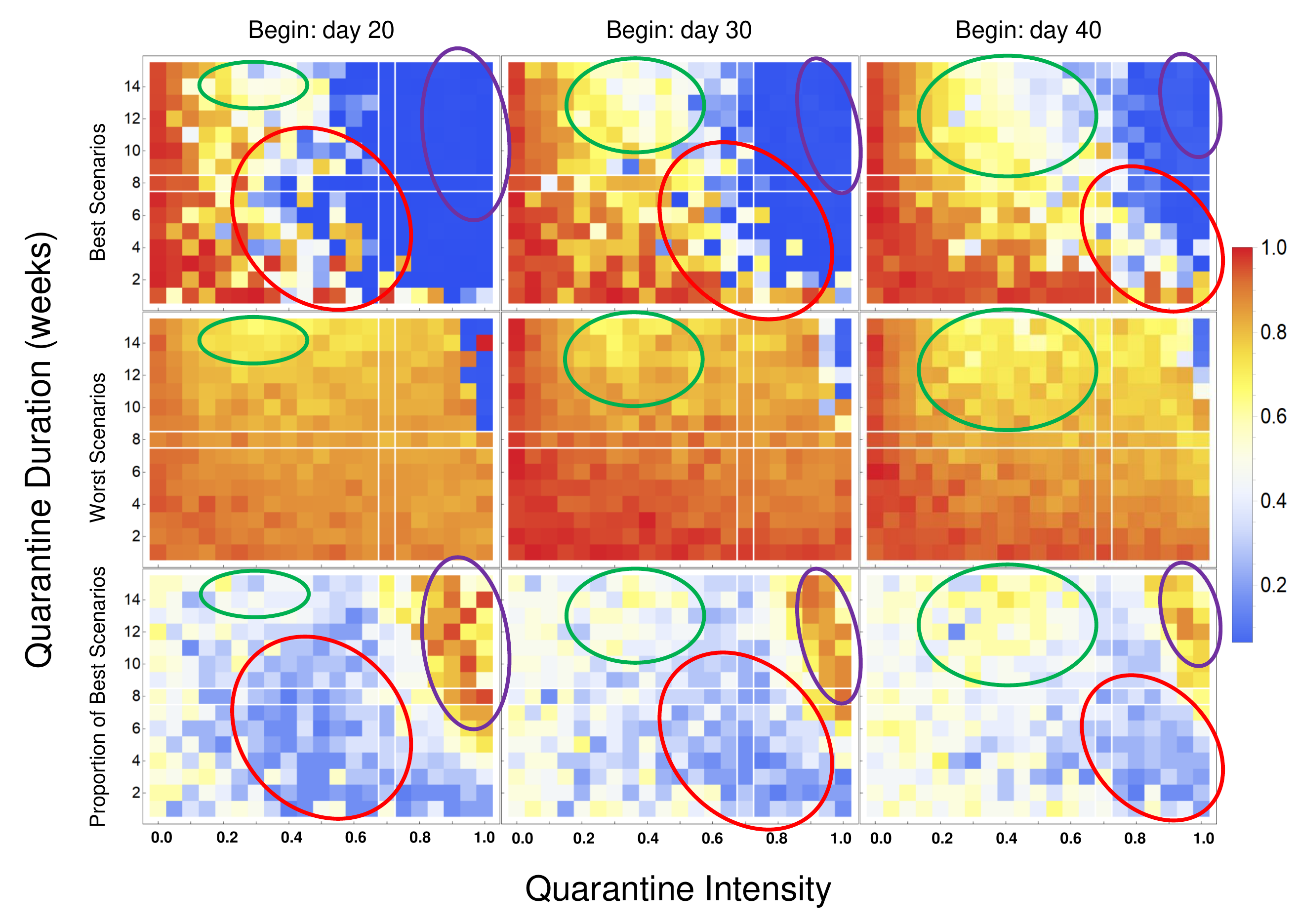}
\end{center}
\caption{Peak height with respect to the average `no quarantine' result, starting 20, 30 or 40 days after the first infection (left, middle and right columns respectively). Plots in the first and second rows show the best and worst scenarios. The third row shows the probability that a simulation results in a best scenario. Quarantine duration is measured in weeks (from 1 to 15) and quarantine intensity goes from 0 (no quarantine) to 1 (full individual lock-down, $p=0$). Green, red and purple ellipses highlight parameter regions of interest. White vertical and horizontal reference lines mark $Q=70\%$ and $t_d=8$ weeks.}
\label{fig:scenarios}
\end{figure}

\begin{figure}[th]
\begin{center}
\includegraphics[width=16cm]{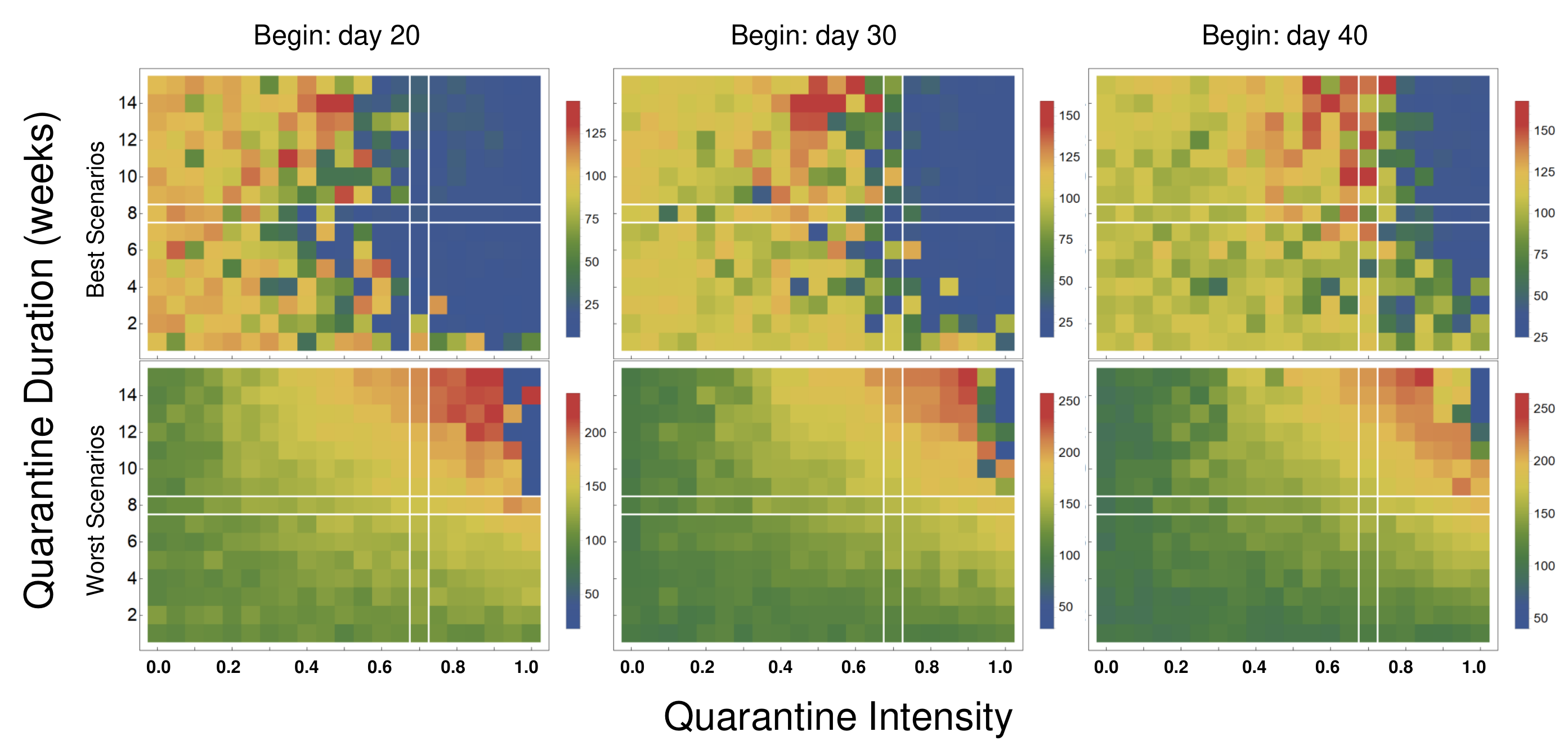}
\end{center}
\caption{Peak center (in days after the first infection) starting 20, 30 or 40 days after the first infection (left, middle and right columns respectively) for different quarantine intensities for best and worst scenarios. }
\label{fig:times}
\end{figure}

\begin{figure}[th]
\begin{center}
\includegraphics[width=16cm]{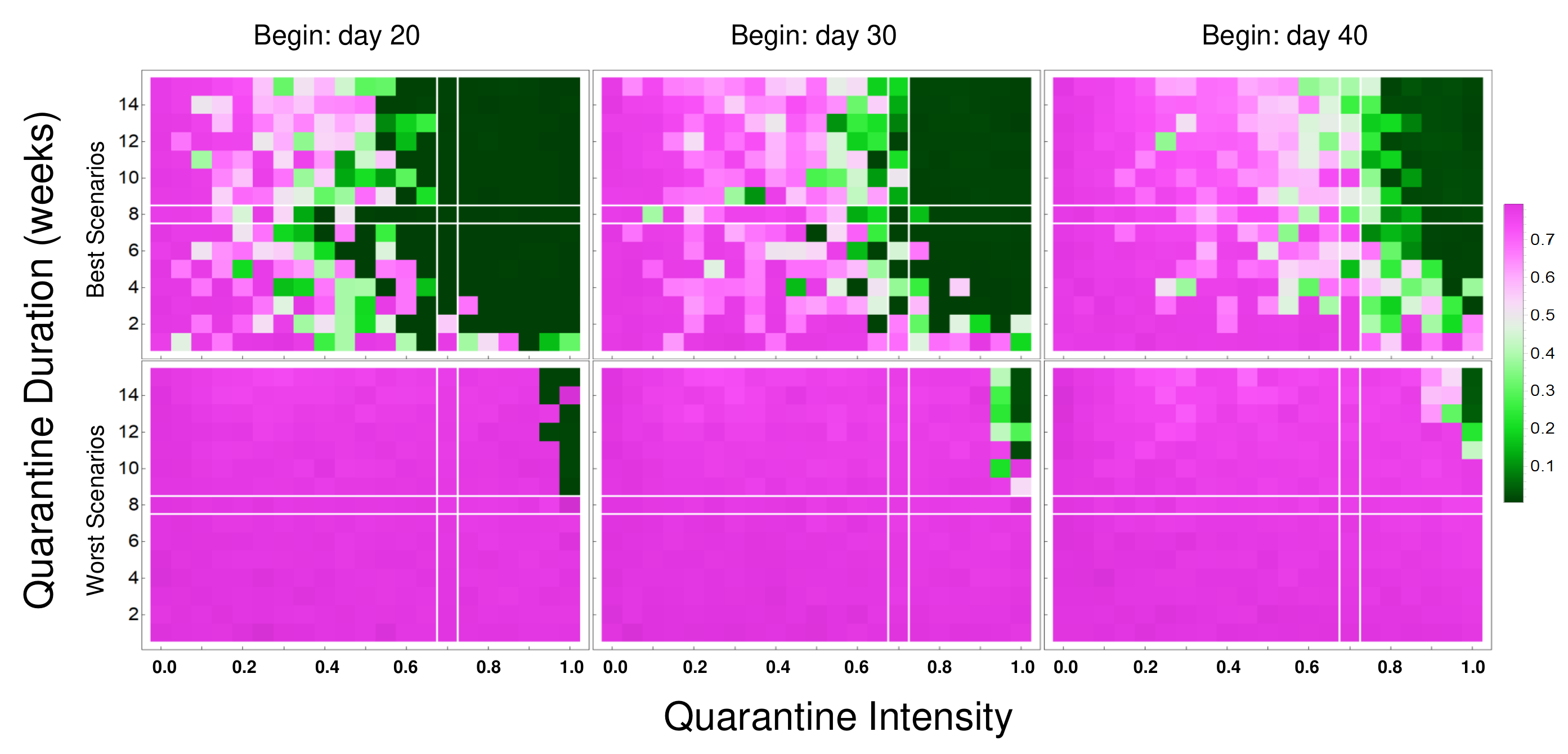}
\end{center}
\caption{Proportion of recovered individuals at the end of the epidemic for quarantine starting 20, 30 or 40 days after the first infection (left, middle and right columns respectively) for different quarantine intensities for best and worst scenarios. }
\label{fig:final}
\end{figure}

Figures \ref{fig:scenarios}, \ref{fig:times} and \ref{fig:final} show results for average peak height, time of infection peak and fraction of recovered individuals at the end of the epidemic (i.e., all individuals that had contact with the virus, as we do not take mortality into account). The results in each case are separated into best and worst scenarios and we compute the probability that a best scenario will happen. For example, a specific set of parameters might result in ending the epidemic, but its probability of occurrence can be too low, excluding it as a recommended policy. All results are displayed as heat-maps.

Fig. \ref{fig:scenarios} shows how peak height varies with quarantine duration, intensity and start date. This information is complemented by Fig. \ref{fig:times}, that shows how peak center changes with quarantine parameters, and Fig. \ref{fig:final}, displaying the proportion of recovered individuals at the end of the epidemic. The purple ellipse in Fig. \ref{fig:scenarios} marks the parameter region where quarantine is very intense and lasts for more than 8 weeks, an ideal situation that works around $90\%$ of the times but is very hard to enforce in practice. In this case the epidemic stops quickly (blue areas in Fig. \ref{fig:times}) and less than $10\%$ of the population is infected (green areas in Fig. \ref{fig:final}).

The red ellipse in Fig. \ref{fig:scenarios} shows a transition zone where the best scenario corresponds to substantial curve flattening. The center of the red ellipse is at $Q \approx 0.5$ for $t_s=20$ but shifts to $Q \approx 0.9$ for $t_s=40$, showing the importance of starting quarantine early. For all values of $t_s$ the red ellipse is centered at $t_d \approx 6$ weeks, which is a relatively short duration. Peak center, however, is not delayed in the best case scenarios. Importantly, best case scenarios are very unlikely in this region, occurring with probability around $20\%$. 

Finally, the region surrounded by the green ellipse in Fig. \ref{fig:scenarios} corresponds to long but moderate intensity quarantines. For the three values of $t_s$ considered peak height was reduced by about $50\%$ in the best case scenarios, which happens about $50\%$ of the times. Peak center was not significantly delayed in the best scenarios, but was pushed forward in the worst scenarios, where peak height was reduced to about $70\%$ with respect to non-quarantine height.  Interestingly, in both scenarios about $70\%$ of the population was infected at the end of the simulation, showing that herd immunity was achieved (corresponding to the pink areas in Fig \ref{fig:final}).

\begin{figure}[th]
\begin{center}
\includegraphics[width=16cm]{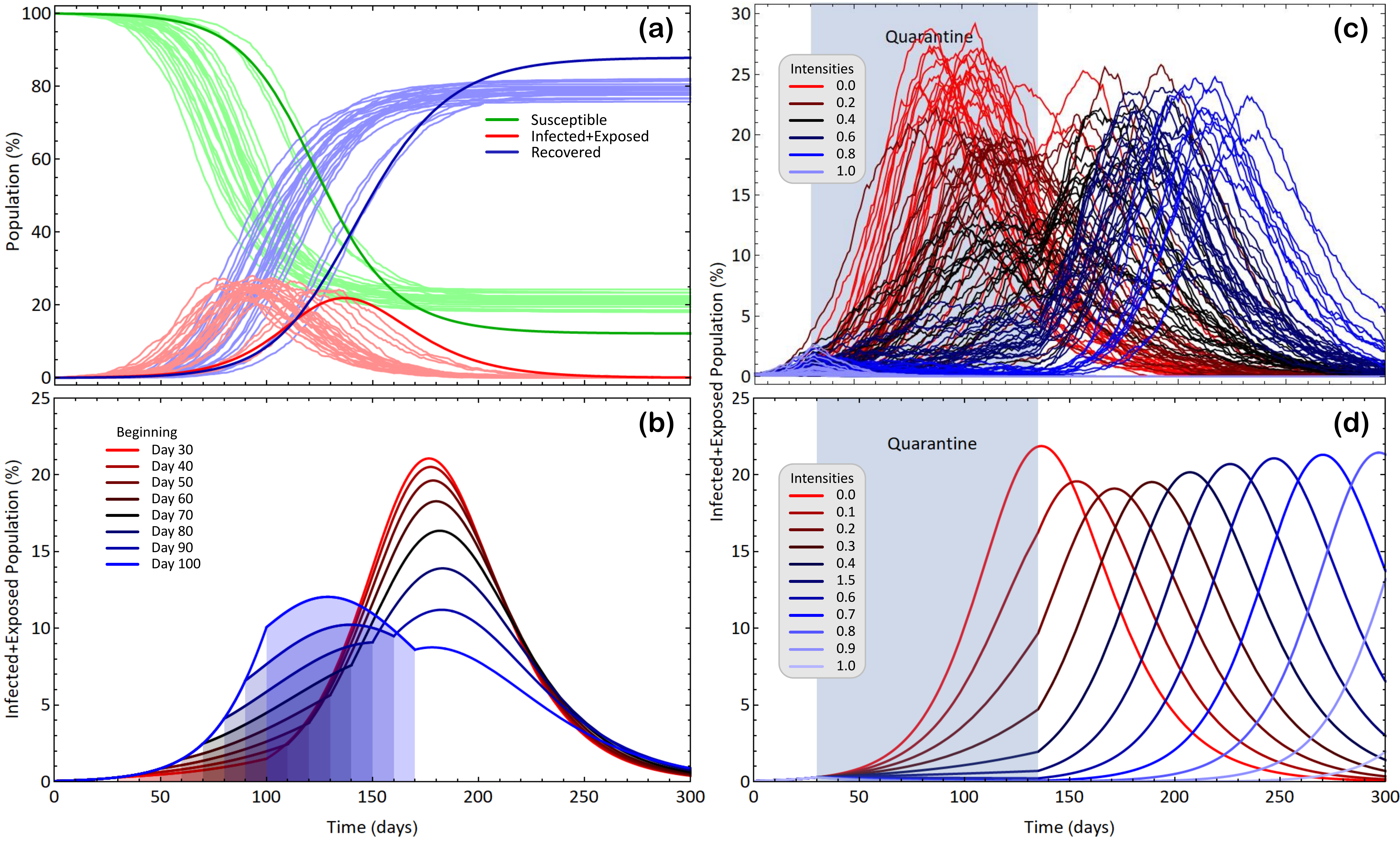}
\end{center}
\caption{(a) dynamics without quarantine computed with mean field equations (thick lines) and IBM simulations (thin lines); (b) mean field results with $Q=0.35$, $t_d=10$ weeks and several starting dates $t_s$; (c) 25 IBM simulations and (d) mean field dynamics for $t_s=30$ days, $t_d=15$ weeks and several intensities $Q$. For the mean field equations, we set $N=2000$, $\beta=R_0\gamma$, $\gamma=1/14$, $\sigma=1/\langle t_i\rangle$, and starting with one infected individual.}
\label{fig:deter}
\end{figure}

Quarantine can also be implemented in the mean field model, Eqs. (\ref{seir}).\cite{morris2020} This is accomplished by integrating the dynamical equations with the infection rate $\beta_0$ for $t \in [0,t_s]$, with the reduced value $\beta_Q = (1-Q) \beta_0$ during quarantine period $t_s < t < ts+ t_d$ and again with $\beta_0$ for $t>t_s+t_d$. Fig. \ref{fig:deter} shows how results of mean field model differ from the IBM simulations. Panel (a) shows the dynamics without quarantine according to the mean field (thick lines) and 25 simulations with the IBM. Panel (b) shows the effects of quarantine on the mean field model for $Q=0.35$, $t_d=10$ weeks and several starting times $t_s$. According to the mean field model quarantine is effective only if started later, otherwise the infection curve peaks at high values when the quarantine is over. The right panels compare IBM simulations (c) and mean field results (d) for $t_s=30$ days and $t_d=15$ weeks for several quarantine intensities $Q$. The mean field infection curves always grow to high values when quarantine is over, whereas the IBM simulations show many examples of low peak values and total epidemic control, with $I+E$ going to zero after the quarantine period. This highlights the importance of heterogeneous social interactions represented by the Barab\'asi-Albert network and stochastic dynamics in epidemiological modeling.

\section{\label{sec:discussion}Conclusions}

In this paper we considered the effects of quarantine duration, starting date and intensity in the outcome of epidemic spreading in a population presenting heterogeneous degrees of connections. The model is stochastic and curves representing numbers of infected individuals vary considerably from one simulation to the other even when all model parameters are fixed. In order to distinguish between different outcomes we have divided them into two groups with the best and worst results based on the height of the infection peak (below or above the average height, respectively). 

We have further divided the results into four qualitative classes delimited by the three ellipses in Fig. \ref{fig:scenarios} plus the rest of the diagram. Besides the obvious region indicated by the purple ellipse where quarantine is very intense and long, we found that short but not so intense quarantine (red ellipse) does not work, since the probability of an outcome in the best scenario is very low. Instead, long but average intensity quarantine is both likely to work and flattens the infection curve by around $50\%$, being the best alternative given the current assumptions.
Indeed, the infection peak is considerably delayed in the region of the green ellipse when it falls into the worst scenario, confirming it as the best bet for preventing the health system breakdown (Fig. \ref{fig:times} ). The proportion of the population that had contact with the virus at the end of the epidemic (number of recovered individuals, Fig. \ref{fig:final}) leads to more than $60\%$ of the population, very close to achieving herd immunity. Comparing to the other regions, this seems to be the best option to control the epidemics under the model assumptions. We note, however, that the model does not account for deaths. If achieving herd immunity implies high mortality, the best option would be long and intense quarantine (purple ellipses in Fig. \ref{fig:scenarios}), the only way to avoid large number of infections and, therefore. high mortality.

We found that differences between mean field and stochastic models are very significant with respect to the effects of quarantine. In many cases as the former cannot control the epidemic, as the infection peak grows again once the quarantine period is over, whereas the latter can end the epidemic in the best case scenarios. 

We recall that we used uniform decrease in infection rate as a proxy for quarantine. This is a simplified approach and other methods could be implemented to verify the robustness of the results. Also, different network topologies might affect the spread of the epidemics. Random uniform (Erdos-Renyi) \cite{albert2002} networks should produce results similar to mean field simulations, but small-world \cite{watts1998,albert2002} or other topologies could speed up or slow down the spread dynamics. 

Our model is particularly suited to study spread between connected cities, that can be represented by modules of a larger network. We have also kept information about the virus DNA and its mutations, allowing us to reconstruct the phylogeny and classify its strains as it propagates. These results will be published in a forthcoming article.

\begin{acknowledgments}
We thank Flavia D. Marquitti for a critical reading of this manuscript. This work was supported by the S\~ao Paulo Research Foundation (FAPESP), grants 2019/13341-7 (VMM), 2019/20271 and 2016/01343-7 (MAMA), and by Conselho Nacional de Desenvolvimento Cient\'ifico e Tecnol\'ogico (CNPq), grant 301082/2019-7 (MAMA). 
\end{acknowledgments}

\end{document}